\documentclass[conference]{IEEEtran}
\pdfoutput=1
\usepackage{graphicx}

\usepackage[bookmarks=false]{hyperref}

\usepackage{xcolor}

\usepackage[english]{babel}
\usepackage{array}
\usepackage{float}
\usepackage{enumitem}
\newcolumntype{P}[1]{>{\centering\arraybackslash}p{#1}}
\usepackage{multirow}
\usepackage{graphicx}
\usepackage{comment}
\usepackage[belowskip=2pt,aboveskip=1pt]{caption}
\usepackage{subcaption}
\begin{document}
%
\title{UNBLOCK: Low Complexity {Transient} Blockage Recovery for Mobile mm-Wave Devices}


\author{\IEEEauthorblockN{Venkata Siva Santosh Ganji, Tzu-Hsiang Lin, Francisco A. Espinal, and P. R. Kumar, Fellow, IEEE}

}

\maketitle
\begin{abstract}
Directional radio beams are used in the mm-Wave band to combat the high path loss. The mm-Wave band also suffers from high penetration losses from drywall, wood, glass, concrete, etc., and also the human body. Hence, as a mobile user moves, the Line of Sight (LoS) path between the mobile and the Base Station (BS) can be blocked by objects interposed in the path, causing loss of the link. A mobile with a lost link will need to be re-acquired as a new user by initial access, a process that can take up to a second, causing disruptions to applications. UNBLOCK is a protocol that allows a mobile to recover from transient blockages, such as those caused by a human hand or another human walking into the line of path or other temporary occlusions by objects, which typically disappear within the order of $100$ ms, without having to go through re-acquisition. UNBLOCK is based on extensive experimentation in office type environments which has shown that while a LoS path is blocked, there typically exists a Non-LoS path, i.e., a reflected path through scatterers, with a loss within about $10$ dB of the LoS path. UNBLOCK proactively keeps such a NLoS path in reserve, to be used when blockage happens, typically without any warning. UNBLOCK uses this NLoS path to maintain time-synchronization with the BS until the blockage disappears, as well as to search for a better NLoS path if available. When the transient blockage disappears, it reestablishes LoS communication at the epochs that have been scheduled by the BS for communication with the mobile. 
UNBLOCK is a low complexity, power-efficient, dynamic blockage recovery protocol. The design is based on extensive measurements and mobility experiments. Experimental evaluation in the variety of office environments available to us shows that UNBLOCK is able to avoid outage and maintain time synchronization in $96\%$ of the transient blockages.
\end{abstract}


%
\IEEEpeerreviewmaketitle

\section{Introduction}
 Next-generation wireless communication technologies (IEEE $802.11$ ay, $5$G and beyond) can enable extremely high throughput applications due to their operation in the mm-Wave spectrum. They promise low latency and ultra-reliable packet delivery. 
To overcome high path loss, 
mm-Wave devices use directional radio beams for communication. 
The device modems typically use small-sized arrays as they offer very high gains and a multitude of radiation patterns.
Deploying a large number of Base Stations (BS) and Access Points can therefore provide Line of Sight (LoS) communication with the mobiles.

However, the mm-Wave spectrum not only suffers from high path loss, but also high penetration loss \cite{penloss, _2017_study}. 
Common media such as drywall, wood, glass, concrete, and the human body, cause severe signal degradation \cite{block1}.
Penetration losses increase 
with the frequency of operation and can completely disrupt LoS communication.
During a blockage event, the Received Signal Strength (RSS) of the mobile reduces drastically and can result in link outage.
If the BS has to reacquire the link, the re-connection to a 5G New Radio (NR) Base Station can take up to a second \cite{_2017_nr}.\footnote{During initial access the mobile performs a spatial scan; it searches for the BS beam with the best RSS. The BS also periodically sweeps through all its beams. After acquisition, both BS and the mobile communicate in scheduled epochs using the beam acquired during the initial access. Sudden onset of blockage results in the mobile losing communication with BS over these epochs. 
}
Such high connection latency severely impacts applications \cite{enable}.
In addition, the network reconnection process is also power hungry. Therefore, it is important to minimize the
need for such re-acquisitions.

In this paper we distinguish between ``transient" blockages and ``permanent" blockages. Transient blockages are defined as blockages which last
on the order of a hundred milliseconds, and which end after the blockage event is over.
Transient blockages can occur as the user walks past obstacles or other users walk across the directional mm-Wave link.
Permanent blockage occurs when the LoS beam between the mobile and the BS is permanently (i.e., for a very long time) blocked, and the only solution for
maintaining a connection is to switch to a LoS beam to a different BS that is not blocked, i.e., handoff. 

The focus of this paper is on sustaining a communication link for control messages during a transient blockage.
Poor RSS during blockage events results in the mobile losing time synchronization with the BS. This results in the mobile missing future scheduled epochs for communication with the BS, during which the BS would have been able to align
its directional beam towards the mobile. The mobile relies on the signals transmitted during such epochs to combat high phase noise in the mm-wave spectrum. 
It is therefore critical to sustain the control plane communication link during such blockage events so as to maintain time-synchronization with the BS.

 The ability to electronically steer the direction of beams can help the device to sustain a communication link by using alternate or Non-Line of Sight (NLoS) paths, when the direct LoS beam is blocked. NLoS paths are reflections from scatterers that exist indoors and urban outdoors. In our extensive experiments in office-like environments, we have found that there typically are available usable NLoS paths between the BS and the mobile. By maintaining the time synchronization and control plane communication with the base station using NLoS paths, mobile devices can sustain a control link through a transient blockage. This then allows recovery to a LoS link as soon as the blockage disappears. 
 {Our experiments} have shown that the RSS on the NLoS paths is about  $10$ dB less than the LoS path in an indoor environment.  This is sufficient for sustaining a NLoS link over which control packets are transmitted.

To employ a NLoS link during blockage events, both the BS and the mobile need a mechanism to identify {the viable} NLoS paths. As the scattering environment changes with user mobility, a one-time environment scan to extract NLoS paths is not sufficient for mobiles. Moreover, such an approach increases the memory usage of the modem. Thus it is necessary to dynamically determine NLoS paths.

The UNBLOCK protocol is a low complex blockage recovery protocol which enables both BS and mobile to dynamically identify NLoS paths, and to then utilize them to maintain time-synchronization so as to preserve the link without need for re-acquisition. UNBLOCK's design is based on extensive mobility experiments using mm-wave software defined radios in indoor {office} environments. It uses an appropriate beam scanning interval that helps in conserving the mobile battery. Unblock employs the right NLoS beam discovery interval for both BS and mobile by observing the duration of transient blockage events, and beam coherence time of NLoS paths from pedestrian mobility experiments. The protocol fits into the framework of $5$G NR standards on which we elaborate in Section \ref{UNBLOCK application in 5G Newradio}.

The rest of the paper is organised as follows. Section \ref{challenges} presents the challenges to overcome to avoid outages to mobiles caused by blockages. Testbed and Experimentation is elaborated in Section \ref{testbed}. Our protocol is described in Section \ref{protocol}. Section \ref{results} provides implementation details and performance of protocol. Section \ref{5gap} maps the protocol  to 5G cellular standards. Section \ref{related}  presents existing work in the domain and Section \ref{conclude} concludes our work.

\section{Challenges}\label{challenges}

Communication devices operating in the mm-Wave spectrum, whether BS or mobile, utilize narrow radio beams to overcome high path loss, and need a beam alignment protocol to handle user mobility \cite{BM}. 
Several in-band and sensor-assisted beam alignment protocols requiring the angle of arrival/departure predict the next best-aligned beam \cite{BA1, BA2,BA3,BA4,BA5}. Although these works address beam misalignment for static links, such predictive mechanisms do not recover link RSS   during LoS blockage events while the user is moving. Maintaining the communication link's RSS during an LoS blockage event perforce requires an NLoS/reflected path.
As the scattering environment changes with user mobility, a one-time environment scan is not sufficient to identify usable reflected paths. 
Such reflected paths can therefore
be discovered only by active probing.

Let $ N_{BS} $ and $ N_{MS}$ be the number of  beams\footnote{In the description of
the following protocol, the BS is using its beams in transmit mode, while the mobile is using its beams in receive mode. By reciprocity, their roles can be revered to obtain the reverse link.}  available at the BS and the mobile respectively. For a given BS transmit beam, a mobile might discover that one or more of these $N_{MS}$ beams (used in receive mode) are NLoS paths. Similarly, for a given receive beam of the mobile, the BS might find one or more of its transmit beams as NLoS paths to the mobile. Both the BS and the mobile must periodically search for and \emph{store} the discovered NLoS paths,  while  communicating on the LoS beams. Storing the discovered paths for later use is necessary since the blockage events are unpredictable. When blockage happens the discovered NLoS beams in the memory are employed to revive link RSS during blockage. Taking the example of a static user where the scattering environment remains the same throughout, for a given LoS transmit beam of the BS, the mobile sweeps through all its $N_{MS}$ receive beams to identify available NLoS beams. Due to the presence of a rich scattering environment indoors, especially due to the walls, the mobile potentially discovers multiple NLoS paths. From our extensive experiments, we found that the RSS of these NLoS beams is typically about $10$ dB less than that of the LoS beams. By listening on the discovered NLoS beam, the mobile can help the BS discover its available NLoS transmit beams {corresponding to the receive beam discovered by the mobile}. To accomplish this, the BS transmits over all the $N_{BS}$ beams while the mobile is making measurements on the NLoS beam. At the end of this process, an NLoS BS-mobile beam pair is found. The BS and the mobile thereby store in their memory at least $N_{BS}$ NLoS beam pairs, one for each BS transmit beam, and use the appropriate stored pair, corresponding to the particular BS transmit beam at the onset of blockage, in case of blockage. Due to mobility, 
the scattering environment changes over time, and so 
the above protocol is periodically employed to discover NLoS beams at appropriate $100$ ms intervals. 

An important additional aspect to consider while designing a protocol for battery-driven mobiles is the reduction of modem power consumption. Frequent measurements using the beams to discover NLoS paths not only consume wireless resources, but persistent usage of the radio front end also consumes a significant amount of device power. To discover an NLoS beam pair, the mobile must first perform a spatial scan, measure RSS, and identify a good NLoS receive beam.
Then it initiates a BS transmit beam sweep to discover a good NLoS beam for the BS. With a large number of beams both at the BS and the mobile, this process requires the mobile to make a significant number of measurements. For example, with a beamwidth of $5^\circ$ to cover a $120^\circ$ sector, at least $24$ beams are needed; therefore the NLoS beam discovery process may require up to $48$ measurements. As the scattering environment changes with mobility, this process must be repeated periodically. The periodicity of this process depends on the Beam Coherence Time (BCT) \cite{BCT}, defined as the duration after which a beam change is necessary to restore the link RSS to the previous maximum. The NLoS beam discovery protocol must run at least once within each BCT on the mobile. The BCT is dependent on motion -- the faster a user is moving, the lesser is the BCT.     

To summarize,  the complete protocol to recover the link RSS during blockage events has the following features:
\begin{itemize}
    \item A dynamic mechanism to discover NLoS beams both at BS and mobile.
    \item Storage of the most appropriate backup NLoS beam pairs.
    \item Optimization of the device power consumption by choosing the optimal NLoS beam pair re-scanning interval.

    \item Most importantly, employing the discovered NLoS beam pair when blockage happens, to avoid the need for network re-connection after transient blockage events.
\end{itemize}

The UNBLOCK protocol's parameters have been tuned based on extensive measurement data and it has been validated in various indoor office environments. We first ran extensive measurements in  Wisenbaker  Engineering Building, a three-storied, 30,000 sq. ft. department building \cite{_aggie}. Experiments have been performed in several classrooms, corridors, and graduate offices to study the RSS of reflected paths from various surfaces. We found that the RSS of the reflected beams is around $10$ to $14$ dB less than that of LoS Beams in these environments. Along with measurement studies, we did experiments to study the BCT using NLoS beams for several pedestrian motion patterns. Based on these observations, we have designed the UNBLOCK protocol to be a low complexity, power-efficient, and dynamic blockage recovery protocol.
\captionsetup[subfigure]{labelformat=empty}
  \begin{figure*}[t]
	\centering
	\begin{subfigure}[t]{.3\linewidth}
		\centering
		\includegraphics[width=2.2in, height=1.5in]{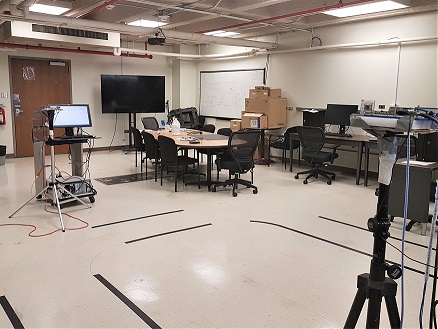}
		\label{fig:b1}	
	\end{subfigure}
	\quad
  \begin{subfigure}[t]{.3\linewidth}
		\centering
		\includegraphics[width=2.2in, height=1.5in]{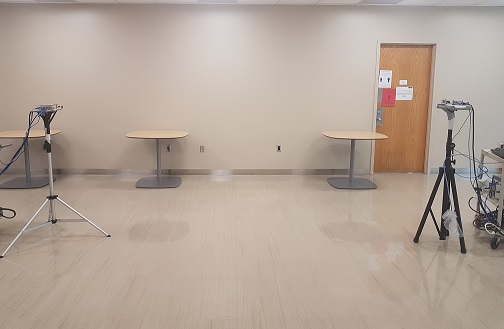}
	
	\end{subfigure}
	\quad
	\begin{subfigure}[t]{.3\linewidth}
		\centering
		\includegraphics[width=2.2in, height=1.5in]{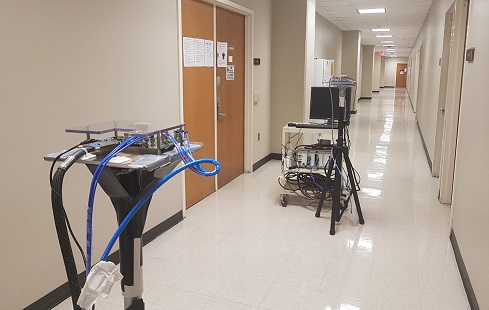}
				\label{blockage}

	\end{subfigure}
		\caption{Indoor Measurement Environments ({From Left to Right: A large lab, a common area, and a corridor)}}
		\label{blockagemeasurement}
\end{figure*}
\section{Testbed and Experimentation}\label{testbed}
The fundamental goal of our experiments is to understand both the signal and temporal characteristics of  multipath/NLoS paths in mm-wave spectrum.
Our experiments have been focused on NLoS path behaviour in indoor environments where  reflected paths are typically present, as our experiments have shown. Accurate understanding of multipath characteristics is important to design a protocol to recover link RSS during blockage events. In crowded indoor environments like conference halls and classrooms, humans act as primary blockers, whereas in household and workplace environments, mobility of the user can result in transient blockage. We conducted experiments in various locations 
of our large department building as a representative of office building environments, using $60$ GHz software defined radio testbed \cite{instruments_2020_introduction}.

\subsection{Testbed}\label{testbedsection}
The testbed is built around a National Instruments' software-defined mm-wave transceiver system with two nodes \cite{instruments_2020_introduction}. 
The node has a chassis with high speed backplane interconnections between several slots. The FPGA cards inserted in these slots communicate with each other through backplace interconnections and are programmed to form transmit and receive chains of mm-Wave radios.  A Sibeam phased array is interfaced with each node. These phased arrays have $24$ antenna elements, $12$ each for transmit
and receive beamforming. With independent transmit and receive RF chains, this provides analog beamforming. Although it is possible to create a large number of beams with arrays, we used $25$ equally spaced narrow beams, roughly covering a sector of $120$ degrees, obtained by programming  $2$ bit weights for each element. The element phase weights are predetermined to obtain the desired beam patterns of approximately $20$ degrees beamwidth. The $25$ predetermined beams form the beamforming codebook. For ease of understanding, we refer to the two bidirectional nodes as ``base station" and ``user".

  In the measurement experiments, we transmit single carrier symbols of $2$ GHz bandwidth at $60$ GHz carrier frequency in a slot that has a duration of $100$ microseconds. A frame of $10$ ms duration is used for time division duplexed transmission. A frame is divided into $100$ slots, with the first $50$ slots for downlink and the rest for the uplink operation. $4$ slots in every $50$ with reference signals are used to time synchronize the base station and user nodes. 
  
  During our measurement studies, the user node measures the RSS and signal-to-noise ratio (SNR) of transmission from the base station node. A beam index from the chosen codebook can be changed every slot; therefore, our codebook which has $25$ different indices can be swept in $25$ slots, i.e., $2.5$ ms. We performed experiments with two different codebooks, covering azimuth and full space. Beams in each codebook approximately cover $120^{\circ}$ sector. Our narrow beam codebook has an approximate beamwidth of $20^{\circ}$.

\subsection{RSS of Reflected paths}

Upon encountering a medium,
electromagnetic radiation may be transmitted, reflected, or absorbed. The reflectivity of radiation from a surface usually increases as the frequency increases. Based on the penetration depth, radiation may transmit through or get completely absorbed by the medium. Because of its shorter wavelength, mm-wave radiation is reflected by common building materials. Penetration and reflection losses \cite{a1_analysis} from our experiments for some of the common obstacles are tabulated in the Table \ref{losses}.

\begin{table}[H]
\small 
    \caption{Penetration and Reflection Losses}\label{losses}
	\centering

		\begin{tabular}{ |P{2 cm}|P{2cm}|P{2cm}|}
\hline		
\hline
    {Material } & {Penetration Loss (dB)} & {Reflection Loss (dB)} \\
 \hline\hline
  Drywall - single layer & 8  &  10 \\ 
 \hline
Drywall - double layer & 16 & 10 \\ 
 \hline

Wooden door & NF & 11 \\ 
 \hline

 Concrete wall & NF & 13 \\ 
  \hline
Human Body & NF & 14-20 \\
 \hline
\hline
\end{tabular}
\end{table}

The reflected radiation from typical building surfaces is measurable and usually above the noise threshold. The total noise power from all the sources in our system at the room temperature for a $2$ GHz bandwidth system is
$-73$ dBm, in the scenarios where RSS is below this noise threshold, we indicated Noise Floor (NF){\footnote{Wireless communication system cannot detect any signal below its NF.}}. 
Therefore we can rely on these reflected paths to maintain the link, perhaps at a reduced rate, during the temporary blockage.

To quantify the RSS of reflected signals, measurements were taken at more than $50$ locations inside the building. In particular, several measurements were taken in office-like environments, corridors, and spaces with concrete walls and dry walls. Fig. \ref{blockagemeasurement} shows some of the test environments. We observed that, on average, the RSS of reflected paths from wooden walls is $10$ dB lesser than that of the LoS path, whereas it is $13$ dB lesser from concrete walls. To make accurate measurements of RSS on the reflected beams, it was ensured that the additional path loss between the reflected path and LoS path was negligible. Through observations, it was found that narrow NLoS beams have better RSS than wider beams, but system designer must also take into account the trade-off between NLoS beam discovery latency and the gains offered for a given beamwidth. As the spatial scan area is lesser for narrow beams, more measurements are required to discover NLoS beams. From the experiments, it was observed that to recover from blockage during mobility, it suffices to keep just one beam,  the best reflected/NLoS beam, in the memory.  

Another important observation from the experiments is that during the blockage events, NLoS beam operation at the mobile alone
can recover RSS and maintain the link in the scenarios where the RSS is above the noise floor of the mobile's receiver. Optimizing both NLoS beams, both at the BS and the mobile, however, can improve the link RSS up to $10$ dB more than that of using NLoS beam alone at the mobile.

\subsection{Beam Coherence Time}
\begin{figure}[h]
  \centering
 \includegraphics[width=1\linewidth, height=1.5in]{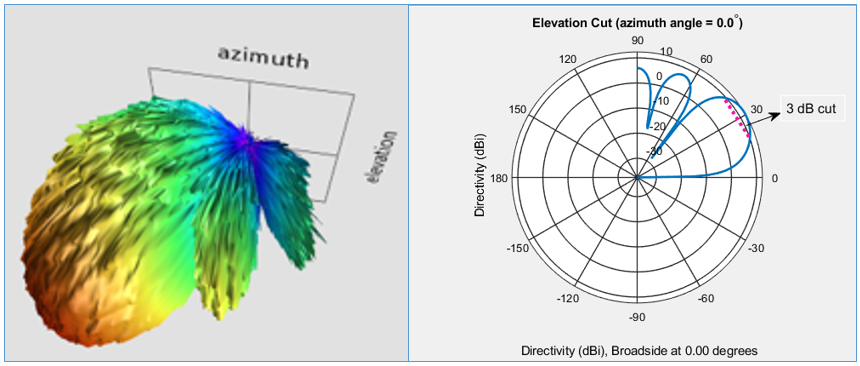}
  \caption{Radiation Pattern }
  \label{bmp}
\end{figure}

It is equally important to study how the RSS of a particular reflected beam changes with user mobility. As the user moves from one location to another, the receive beam of the mobile has to be changed to better align with that of the BS beam. As the alignment is disrupted over time, the RSS of the beam is degraded. How long a particular mobile beam remains aligned with that of the BS depends on the mobility of the user. As we rely on scatterers for blockage recovery, which are more common in indoor rather than outdoor environments, the focus was on pedestrian mobility. The temporal changes in RSS for both translational and rotational motion patterns of the user were studied. For any beam pattern, the high gain region is within  $3$ dB  from the maximum or between the  half power points of the main lobe. Beyond the the $3$ dB beamwidth, the gain in the angular direction falls off significantly from the peak and is often unpredictable. Fig. \ref{bmp}  
shows the radiation pattern of the beams used in the experiments. Therefore, it is necessary to change the beam after misalignment results in  more than $3$ dB loss. We formally define this duration as the ``Beam Coherence Time" \cite{bct1, bct2}. It is the duration after which the RSS on a beam drops by $3$ dB from the maximum. 

Similar to the measurement experiments, mobility experiments were performed at several locations in the building. Different mobility patterns like human walk, hand movements, and array rotation at multiple angular speeds, were studied. Knowledge of BCT helps in determining the NLoS beam discovery interval. Choosing the right interval for beam discovery not only helps in reliably preserving communication during the blockage but also reduces the device power consumption. Table \ref{BCT}a presents the BCT for users walking at distances of $5$ m and $10$ m from the BS, with the mobile using a fixed NLoS beam to communicate with the BS. Table \ref{BCT}b shows the BCT for multiple angular speeds using a NLoS beam,  when the distance between BS and mobile is $5$ m. 
\begin{table}[h]	
	\centering
	\small 
	\caption{Mean BCT}\label{BCT}
	\begin{subtable}[t]{1.5in}
		\centering
		\begin{tabular}{ |p{.75cm}|p{1.1cm}|p{.75cm}| }
\hline
 \hline
 Speed (m/s)  & Distance (m) & BCT (ms)\\ 
 \hline\hline
  .8 & 5 & 600  \\ 
 .8 & 10 & 800  \\ 
  1.4 & 5 & 470  \\ 
 1.4 & 10 & 750  \\ 

 \hline
\hline
\end{tabular}
		\caption{Translational Motion}\label{table:1a}
	\end{subtable}
	\quad
	\begin{subtable}[t]{1.5in}
		\centering
		\begin{tabular}{ |p{.75cm}|p{1.1cm}|p{.75cm}| }
\hline
 \hline
 Speed (rad/s)  & Distance (m) & BCT (ms)\\ 
 \hline\hline
  2$\pi$/9 & 5 & 284  \\ 
 $\pi$/3 & 5 & 200  \\ 
 2$\pi$/3 & 5 & 141  \\ 
 4$\pi$/3 & 5 & 101  \\ 
 \hline
\hline
\end{tabular}
		\caption{Rotational Motion}\label{table:1b}
	\end{subtable}
\label{table2}
\end{table}
The fundamental idea behind using BCT to determine the NLoS the beam re-scanning interval is that there is likely to be a LoS beam update after this duration, either at BS or user, to overcome the misalignment caused by the user mobility. If the LoS beam either at the BS or user is changed, then the previously discovered NLoS beam is no longer  valid to recover from blockage, and must be updated to account for  mobility of the user. From the experiments, it was observed that rotational motion of the user quickly degrades the RSS, and hence the BCT is lesser compared to lateral user motion patterns like walking. Also, use of narrow beams demand frequent updates of NLoS beams, just as it does for LoS beams. It is expected that rapid hand movements while users are interacting  with the mobile result in faster angular speeds, and so have shorter BCT duration. This can especially happen while users are playing mobile games or running Virtual Reality (VR) applications. Other movements, such as swinging hands, head roll, and body rotation, although slower, can cause intermittent  blockage. 

Unlike RSS improvement with NLoS beam operation at the BS in conjunction with NLoS beam at the user, the BCT remains unchanged, as it is purely dependent on the motion pattern of the user. Additionally, the BCT duration becomes larger when the user moves farther from the BS because of the smaller subtended angle.

\section{UNBLOCK}\label{protocol}
Before describing the protocol for blockage recovery, we elaborate on the NLoS beam discovery procedure both at the BS and the mobile. First, the mobile, while listening to a  BS beam using its LoS beam, performs a spatial scan during the available communication epochs with the BS. The mobile measures the RSS on all its beams for a particular BS LoS transmit beam, and discovers NLoS beams with RSS  within   $10$ dB of its LoS Beam. At the end of this procedure, the mobile has obtained an NLoS path to the BS. Next the BS needs to optimize its NLoS beam; this can be achieved by the mobile measuring its RSS using the discovered NLoS beam, while the BS is sweeping over its transmit beams. The mobile communicates back to the BS the identity of the transmit beam that has the highest RSS. Thereby, both BS and mobile discover their respective NLoS beams that can be used in case of blockage.

\begin{figure}[h]
  \centering
 \includegraphics[width=.8\linewidth]{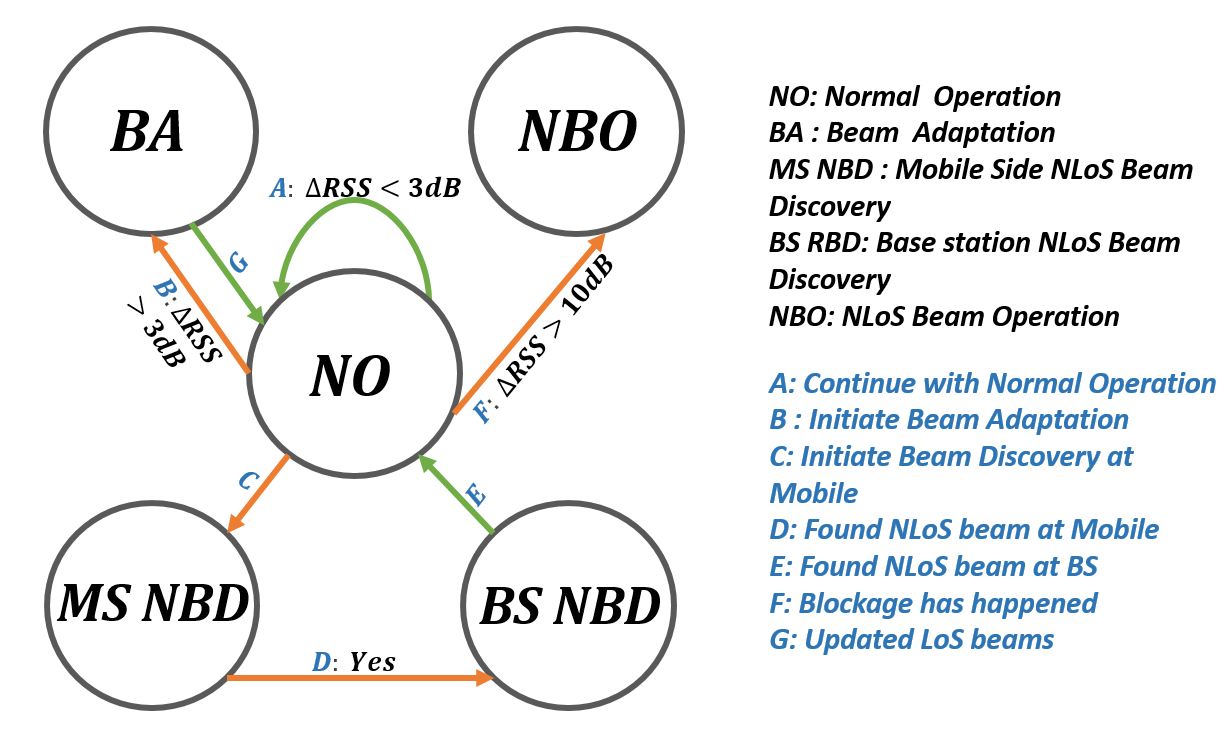}
  \caption{UNBLOCK State Machine }
  \label{statemachine}
\end{figure}

Based on the BCT data in Table \ref{BCT}, the {smallest} BCT is around $100$ ms. Therefore, the NLoS beam discovery procedure is repeated every $100$ milliseconds. Fig. \ref{statemachine} shows the State machine of the UNBLOCK protocol. 
\begin{figure*}[t]
  \centering
 \includegraphics[width=\textwidth]{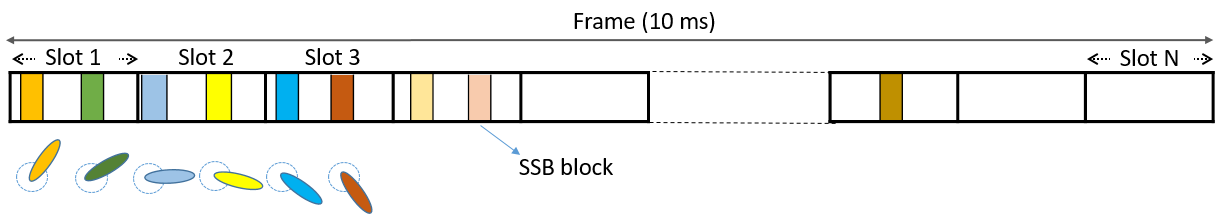}
  \caption{5G NR Frame}
  \label{5gframe}
\end{figure*}
To limit the discussion to the scope of this work, we assume that both BS and mobile run a beam alignment protocol to align their LoS beams to account for user mobility. This happens in the State of ``Beam Adaptation" (BA). The ``NO" state denotes Normal Operation, in which both the mobile and BS continue to use an aligned LoS beam pair. These LoS beams are acquired during the initial acquisition. By the end of the initial acquisition process, both BS and mobile determine their respective LoS beams. Over time, this alignment gets disturbed because of user mobility. Whenever the mobile finds that its current RSS has dropped by $3$ dB from the previous measurement occasion, it moves to the ``Beam Adaptation" (BA) state. In the BA state, the mobile first tries to recover the RSS by aligning its LoS beam, and, later, the BS adapts the LoS beam on an as-needed basis. Upon successful alignment, they continue in the ``NO" state.

Periodically, in our case, every $100$ ms, the mobile moves to the ``MS NBD" state, i.e, ``Mobile Side NLoS Beam Discovery. In this state, the mobile performs a spatial scan to find an NLoS receive beam for the current LoS beam of the BS. The mobile measures the RSS on each of its beams and identifies beams with RSS within 10 dB of the current LoS beam. Among all the NLoS beams discovered, it stores the beam index with the highest RSS. While the mobile is listening to the BS on the NLoS beam, ``Base Station NLoS Beam Discovery" (BS NBD) is triggered. In ``BS NBD", the BS sweeps over all beams, one at a time, and the mobile measures the RSS of the discovered NLoS beam. Upon the mobile finding the BS beam with the highest RSS, it communicates the beam information to BS. The BS stores the NLoS beam information. This process is repeated every $100$ ms to account for changing scatterers caused by user mobility.

If at anytime during ``NO" operation, the current RSS of the mobile drops by $10$ dB from that of the previous measurement, the mobile first moves to the ``NBO" state. In this ``NLoS Beam Operation" state, the mobile first reverts to the stored NLoS beam and informs the BS via the control packet communication to switch the BS beam to the NLoS beam. Upon detecting the blockage event from the control packets, the BS switches to the NLoS beam, retrieving the stored information from its memory. There are ample measurement opportunities in the 3GPP NR standards to discover NLoS beams. Section \ref{UNBLOCK application in 5G Newradio} presents how UNBLOCK can run in adherence to the existing standards.

\section{Implementation and Evaluation} \label{results}
    We implemented the UNBLOCK protocol on a $60$ GHz software defined radio tested. In the testbed, a beam can be changed every $100$ $\mu$s. With $25$ beams in the beamforming codebook, it takes $2.5$ ms each at mobile and BS to search over all the beams. Every $100$ ms, the UNBLOCK protocol on the testbed needs $5$ ms to identify an NLoS backup beam pair. The Beam Alignment protocol proposed in \cite{ganji_10_beamsurfer} for LoS Beam Adaptation was implemented to evaluate UNBLOCK under user mobility. The UNBLOCK protocol was programmed in labVIEW software both at the BS and the mobile nodes. UNBLOCK was evaluated in two different scenarios.
    \begin{itemize}
        \item BS and mobile are stationary, while a human walks around and blocks the LoS beam pair.
        \item BS and blockers are stationary, while a user with a phased array walks in such a way that it obstructs the LoS path to the BS.
    \end{itemize}
     The experiments were repeated in several locations in the building. It is observed that in both scenarios, UNBLOCK is successful in recovering the link during blockage, and thereby preserving the time synchronization. 
     
\begin{figure}[h]
  \centering
 \includegraphics[width=.8\linewidth]{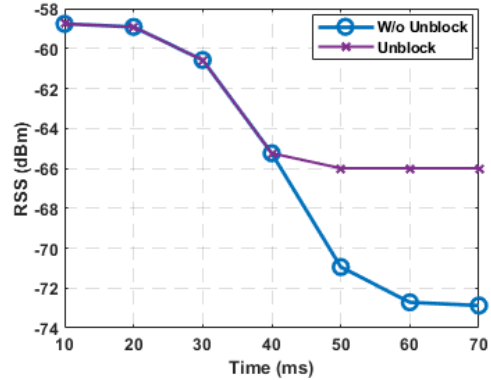}
  \caption{Blockage Recovery}
  \label{result1}
\end{figure}

Fig. \ref{result1} shows a blockage event in an experiment in which a human walks across the path between a static BS and a mobile and briefly blocks the link. It can be observed that the blockage occurs quickly; therefore, it is not possible to predict such events. In Fig. \ref{result1}, the RSS of the LoS link RSS, which is $-58$ dBm, falls to -$72$ dBm in $35$ milliseconds. The noise floor, which is the total noise power from all the sources in our transceiver is $-74$ dBm
  system with $2$ GHz bandwidth is . In the communication system with $2$ GHz operating bandwidth, the thermal noise at room temperature alone is $-80.8$ dBm. When the RSS hits NF, the mobile loses time synchronization and connectivity with the BS. Without the UNBLOCK protocol, one can observe that the mobile would lose synchronization with the BS and will need to re-initiate the initial search process. However, when the UNBLOCK protocol is running, it detects the RSS drop at around $50$ ms,  and the backup beam pair in the memory maintains link RSS to preserve time synchronization. The experiments were repeated at $50$ different locations with the BS remaining static, while the user with the phased array in hand either turns away from the BS to obstruct the LoS links with his body, or walks past a metal board that blocks LoS link. It was observed that the UNBLOCK protocol preserves time synchronization $96\%$ of the experiments.
\section{UNBLOCK for 5G NR } \label{5gap}

 \label{UNBLOCK application in 5G Newradio} \label{UNBLOCK application in 5G Newradio}
Beam Management at the base station, called gNodeB, and user equipment (UE) in $5$G Newradio standards can be broadly divided into two phases: 

\begin{itemize}
    \item Beam Selection phase. During this phase, a mobile performs Acquisition/Re-Acquisition.
    \item Beam Refinement phase. This allows both the BS and mobile to perform Receive/Transmit Beam Adaptation periodically, determine a good reflected beam, and perform blockage recovery upon onset of blockage.
\end{itemize}

The 3GPP standards \cite{_2017_nr} make provisions to complete these two phases at gNodeB and UE to identify an aligned beam pair.
We describe how UNBLOCK can be implemented using the mechanisms available in the standards for these phases.

The gNodeB transmits  synchronization signal blocks (SSB), control information, and broadcast information necessary for UEs to discover and connect to it. This is done with up to $64$ different beams within a sector, every two frames.   Communication between gNodeB and several UEs happens within a frame, which is of $10$ ms duration.  Each frame is further divided into slots. Based on the configuration of deployment, the number of slots in a frame can be chosen from the set $\{10, 20, 40, 80, 160\}$. 
mm-Wave 5G standards are time-duplexed, meaning the gNodeB and the UE communicate in a synchronized order, but not simultaneously. In a slot, either downlink/uplink data or control information
 are transmitted. Within each slot, each SSB is transmitted on a particular beam, and there can be up to $4$ SSB blocks within a slot. Fig. \ref{5gframe} shows a $5G$ NR frame.
 
 When UE is turned on, it first searches for these SSB beams and time synchronizes with  gNodeB. As there is no prior synchronization with gNodeB, the schedule of these SSBs is unknown and UE has to wait at least $20$ ms on a receive beam. As there are $64$ SSBs, 
 it takes $1.28$ seconds to test all the gNodeB beams. 
 
 However, upon successfully connecting to gNodeB, UE is time-synchronized with gNodeB, and is also aware of the SSB beams' temporal locations. In addition, the gNodeB changes their periodicity to $5$ ms from $20$ ms for the particular UE after initial access procedure. Hence, UE has ample opportunity to harvest the reflected beams of the transmitting/gNodeB beam every $100$ ms.  

\subsection{Beam Selection in 5G Newradio} \label{Beam Selection in 5G Newradio}
As the first step in the network connection procedure, a user device operating in the mm-wave spectrum
initiates gNodeB discovery by searching for synchronization signals. It is a directional search using several beams to identify the direction of the best possible RSS. The gNodeB sweeps all $64$ SSBs consecutively, and repeats the sweep every $20$ ms.  UE discovers either one or multiple SSB beams and uses one of those beams to communicate back to gNodeB to complete
the connection. gNodeB changes the  periodicity of these SSB bursts to $5$ ms for UEs after acquistion. Therefore, using the UNBLOCK protocol, it has enough opportunities not only to make measurements on the current receive beam but also to identify the reflected NLoS beams of a transmitting beam of the gNodeB.  
Upon listening to UE's connection request on one of its SSB beams,  gNodeB designates that SSB beam as a communicating beam to that UE. By the end of this procedure, both  gNodeB and UE identify the transmit-receive beam pair for communication.  UE receives all subsequent SSB, channel state information reference signals (CSI-RS) and scheduling information once it connects to the gNodeB. 

 
\subsection{Beam Refinement in 5G Newradio} \label{Beam Refinement in 5G Newradio}
After Beam Selection, the gNodeB assigns each user a set of beams with either SSB or CSI-RS resources for channel quality reporting. 
This helps the gNodeB manage its beam  as and when necessary,
since user motion or hand movements quickly disrupt beam alignment.  In the beam refinement phase, the gNode and UE adapt the transmit-receive configuration identified during the initial connection to handle mobility and to improve the quality of service. 
UE also reports channel state information to gNodeB, which gNodeB uses to adapt its transmit beams. 
As the user moves, the previously identified beams might become stale, either through misalignment or blockage, and new beams are necessary to sustain connectivity. When beam misalignment is detected either at gNodeB through reports
from UE, or at UE from measurements, any beam adaptation protocol can be employed to manage the Transmit and Receive beams. 

UNBLOCK at the mobile side can identify NLoS beams for a given gNode B transmit beam either using SSB resources or CSI-RS signals scheduled on that beam. To find the NLoS transmit beam of gNode, UNBLOCK can use the resources available either for Beam Selection or Beam Refinement. In the case of the Beam Selection phase, gNode B sweeps over all its $64$ transmit beams in $5$ ms. UNBLOCK can also identify the NLoS gNodeB transmit beam from an exclusive set of beams allocated by the gNodeB. It is also possible to use beams in both Beam Selection phase and beam refinement phase to quickly narrow down the NLoS beams. Unlike our implementation, beams can be changed on a per symbol basis. Symbols within SSB are of either $4.46$ or $8.92$ microseconds duration, and CSI RS duration can be $8.92$ microseconds. Therefore, the mobile can use a large number of narrow beams to harvest reflected paths.  Although faster search methods like hierarchical search can be employed, to present the fundamental idea of UNBLOCK we limit the discussion to an exhaustive search.

\section{Related Work} \label{related}
The authors of \cite{blockage_study} have proposed a measurement system and conducted human blockage experiments using a similar testbed \cite{instruments_2020_introduction}. Our experiments and the NYU measurement campaign \cite{blockage_study} confirm the existence of rich multipaths in indoor environments at mm-wave frequencies. The work \cite{adapoon} modelled the common causes of packet decoding errors in mm-wave communications as a linear dynamical system and proposed tests to identify the cause of an error. It identifies blockage after its occurrence, and is a reactive solution unlike the pro-active UNBLOCK protocol which always maintains a backup beam pair. Using multihop mm-Wave network to avoid blockage by forwarding packets to the destination via unblocked links is studied in the work \cite{8580773}.
In the paper \cite{antenna_diversity}, hybrid beamforming is explored to instantaneously recover from blockage. The work proposes antenna diversity to design and maintain several beams in different directions at the transmitter and receiver. In case an antenna beam suffers from blockage, the proposed method invokes one of the beams from the available. Despite requiring simultaneously active beams, the method still needs a search to identify the unblocked beams.  Another reactive approach to overcome blockage is presented in \cite{8589087}. Once blockage occurs, several beams are used simultaneously in this approach \cite{8580773}. Most of the works in the literature propose reactive mechanisms to overcome blockage; in contrast, the UNBLOCK protocol is a pro-active dynamic approach to avoid outage during user mobility and always keep a backup beam pair to preserve time synchronization for immediate use when blockage happens.
\section{Conclusion} \label{conclude}
mm-Wave signals are highly susceptible to blockage. Avoiding link outage and preserving time synchronization during blockage events is important to conserving mobile device power and circumventing high reconnection latency. The UNBLOCK protocol harvests viable
NLoS paths in a timely fashion and immediately employs them upon the onset of blockage. Thereby, it ensures that time synchronization with the BS for mobile devices is maintained during transient blockage events. When the transient blockage disappears, the LoS link is immediately restored by communicating
at those epochs dedicated by the BS for the particular mobile. Measurement experiments and evaluation on software defined radios are reported to assess the efficacy of the UNBLOCK protocol.



%
\bibliographystyle{IEEEtran}
\bibliography{IEEEabrv,main}
\end{document}